\documentclass[a4paper]{jpconf}
\usepackage{graphicx}
\usepackage[square,sort&compress]{natbib}
\newcommand{\sdss}{\emph{SDSS }}

\newcommand{\teff}{$T_{\rm eff}$}

\newcommand{\msun}{$M_\odot$}
\newcommand{\kms}{\ensuremath{kms^{-1}}}

\begin{document}
\title{The HYPERMUCHFUSS Campaign -\newline an undiscovered high velocity population}

\author{Alfred Tillich$^{1}$, Stephan Geier$^{1}$, Uli Heber$^{1}$, Heiko Hirsch$^{1}$, Pierre Maxted$^{2}$, Boris Gaensicke$^{3}$, Tom Marsh$^{3}$, Ralf Napiwotzki$^{4}$, Roy \O stensen$^{5}$ and Chris Copperwheat$^{3}$}

\address{$^{1}$ Dr. Remeis-Sternwarte , Sternwartstr.7, 96049 Bamberg, Germany}
\address{$^{2}$ Astrophysics Group, School of Physics and Geographical Sciences, Lennard-Jones Laboratories, Keele University, ST5 5BG, United Kingdom}
\address{$^{3}$ Department of Physics, University of Warwick, Coventry CV4 7AL, United Kingdom}
\address{$^{4}$ Centre of Astrophysics Research, University of Hertfordshire, College Lane, Hatfield AL10 9AB, UK}
\address{$^{5}$ Instituut voor Sterrenkunde, Katholieke Universiteit Leuven, Celestijnenlaan 200D, 3001 Leuven, Belgium}

\ead{Alfred.Tillich@sternwarte.uni-erlangen.de}

\begin{abstract}
We present an overview and a status report of HYPERMUCHFUSS (HYPER velocity or Massive Unseen
Companions of Hot Faint Underluminious Stars Survey) aiming at the detection of a population of high velocity
subluminous B stars and white dwarfs. The first class of targets consists of hot subdwarf binaries with massive compact companions, which are expected to show huge radial velocity variations. 
The second class is formed by the recently discovered Hyper-velocity stars, which are moving so fast that the
dynamical ejection by a supermassive black hole seems to be the only explanation for their origin. 
Until now only one old HVS has been found, but we expect a larger population. We applied an effecient selection technique for hot subdwarfs and white dwarfs with high galactic restframe velocities from the
\sdss spectral data base, which serve as first epoch observations for our campaign with the ESO VLT and NTT in
Chile, the 3.5\,m telescope at DSAZ observatory (Calar Alto) in Spain and the WHT on La Palma. 
The survey is nearing completion and provides us with promising candidates which will be followed up to measure their RV-curves to uncover massive companions or prove their nature as HVS.
\end{abstract}

\section{Introduction}
Our HYPERMUCHFUSS campaign aims basically at two different fast populations amongst hot subdwarf stars. 
Hot subdwarf stars (sdB/sdOs) are considered to be helium 
core burning stars with very thin hydrogen envelopes and masses 
around $0.5\,M_{\rm \odot}$.
A large fraction ($40-70\%$) of the sdBs are members of short period binaries ($P=0.1-10\,{\rm d}$) 
\cite{2001MNRAS.326.1391M}. They are the result of one or two common-envelope ejection phases in 
which the companion to a giant star is engulfed in the outer layers. Due to dynamical friction the companion 
spirals in towards the core of the red giant and the envelope is ejected. 
About 10\% of these companions are low-mass M stars, the remainder are compact stars, mostly white dwarfs.
In addition 20\% of sdB stars are in wide binaries with G-K type companions.
\begin{itemize}

 \item Geier et al. (this conference) presented the discovery of a hidden population of heavy companions (including stellar black holes) in low inclination hot subdwarf systems. The lack of high inclination systems remains mysteriuos and calls for a dedicated short-period binary survey.
Therefore the first group of our fast targets are short period binaries. 
Geier et al. studied in detail 
50 sdB binaries and derived 
companion masses by measuring the projected rotational velocities and surface
gravities from high resolution UVES spectra. 
For short period systems (P=0.1 to 0.6d) we assume that the stars are 
rotating synchronously with the orbit. 
This assumption has been verified from ellipsoidal light variations 
and asteroseismology. Hence the
method is applicable for orbital periods of up to at least 0.6d, which is 
the case for more than half of the sample. 
While most masses are compatible with normal WDs, a couple of systems have 
unusually heavy companions. 
About 4-8\%  of massive companions were found to exceed the Chandrasekhar mass limit. 
A massive main sequence companion can be excluded as it would outshine the sdB. The companions are therefore most likely neutron stars or black holes. 
All of these systems are of low inclination and have short orbital periods. 
Obviously the inclination does not have a preferred direction and therefore 
also a population of high inclination systems has to exist. 
These systems would then have radial velocity (RV) variations as high as 500 \kms\ at orbital periods of less than 0.5 days. 
Considering the fraction of $4-8 \%$ together with the very low number of known neutron star 
or black hole binaries which are detected as X-ray sources, a confirmation 
would mean a tremendous rise of the total number of these heavy compact objects in the galaxy. 

 \item The second group of our fast targets consists of single stars. Hyper-velocity stars (HVS) have been found among young late B-type stars \cite{2005ApJ...622L..33B} as well as among old population hot subdwarf stars \cite{2005A&A...444L..61H}. The first systematic survey targeted late B-type HVS and has been performed with great success \cite{2006ApJ...640L..35B}. Therefore we initiated the first intensive HVS survey dedicated to old population stars. HVS are predicted to have been formed through the tidal disruption of a binary by the supermassive blck hole (SMBH) in the galactic center (GC)\cite{1988Natur.331..687H}. Other formation scenarious like e.g. the scattering of multiple stars (or binary stars) 
by SMBH+IMBH systems and the multiple star scattering in dense clusters have been proposed recently. This rises the number of potential HVSs significantly and justifies extended searches for these interesting objects. Especially as their trajectories contain information on the important parts of the mass distribution in our galaxy: the GC and the dark matter Galactic potential. 
We focus particularly on the old population HVS stars i.e. hot subdwarf stars as we discovered and successfully analysed the first one of this kind \cite{2005A&A...444L..61H}.

 \item By combining the two parts of the project we might even find a \textbf{\textit{HVS binary}}. Our campaign is a very promising approach for finding such systems, as we target subdwarfs and especially short period binaries. These systems more likely might remain intact after interacting with a SMBH/IMBH pair than others \cite{2008arXiv0810.3848S}. As pointed out by Lu, Yu \& Lin \cite{2007ApJ...666L..89L}, the detection of just one such object would provide strong evidence that the massive black hole in the GC is a binary.
\end{itemize}

\section{Hyper-velocity stars}
HVSs have attracted great interest as they allow implication on a wide field of astrophysics, which we intend to outline here. 

\begin{figure}[t]
\begin{center}
\includegraphics[scale=0.35]{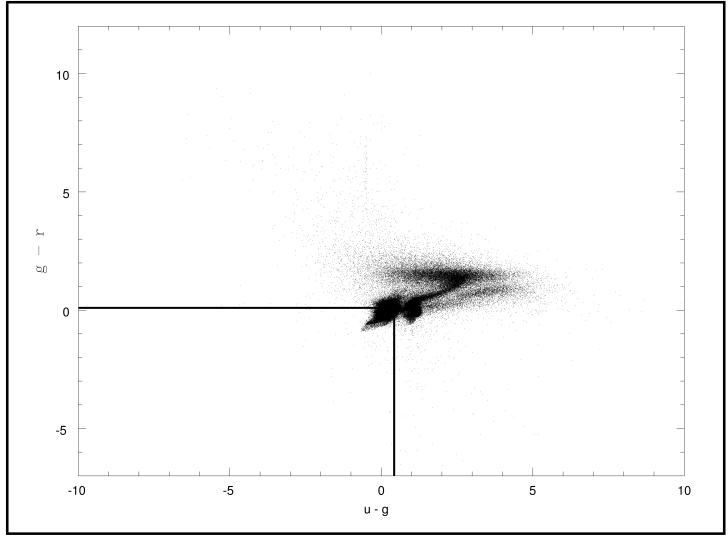}
\caption{\label{label}Colour indices of all stars in \sdss DR6 with the colour criteria used.}
\end{center}
\end{figure}

HVS move so fast that dynamical ejection by massive black holes appears to be the only viable mechanism for their origin. It is well established that the GC hosts a SMBH \cite{2002Natur.419..694S}, \cite{2005ApJ...620..744G}) with a mass of about $3\times10^{6}$\msun.
Already in the late eighties it was predicted from numerical experiments that a star can be ejected from the GC with velocities exceeding the escape velocity of the Galaxy by the disruption of a binary through tidal interaction with a  central massive black hole. \cite{1988Natur.331..687H}.
The detection of one such HVS would be evidence for the existence of a massive black hole in the GC since none of the alternative scenarios put forward to explain the phenomenon of run-away stars with such high velocities \cite{1988Natur.331..687H}. 

Due to the exceptional origin of HVS, they must urgently be separated from the already well known {\it runaway stars}, which are high velocity O and B type stars at high Galactic latitudes with galactic restframe (GRF) velocities in the range of $100-200$\,\kms. 
For {\it runaway stars} we know two capable ejection mechanisms. 
(i) In the binary supernova scenario a massive star undergoes a core collapse explosion and its companion star is released at about its orbital velocity (up to 200\,\kms).
(ii) Close single star/binary (or binary/binary) encounters in open clusters can lead to ejection velocities up to 300\,\kms.\newline
These scenarios have met with considerable success to explain most of the observed run-away stars known \cite{1992ApJ...400..273C}. Recently, the unbound star HD~271791 was found to be an extreme SN binary run-away B-star and termed hyper-run-away star\cite{2008ApJ...684L.103P}. Therefore we suggest a clear classification of the fastest stars. 


The first HVS was found by Brown et al. \cite{2005ApJ...622L..33B}, a faint B-type star with a heliocentric RV of 853\,km/s. Soon thereafter two members of our group (Hirsch et al. \cite{2005A&A...444L..61H} and Edelmann et al. \cite{2005ApJ...634L.181E}) discovered HVSs No.~2 and~3 (a sdO star US~708, $RV=708$\,\kms\, and an early B-main sequence star HE~0437$-$5439, $RV=723$\,\kms), both unbound to the Galaxy.
Dedicated systematic surveys for late B-type stars have resulted in the discovery of additional 13 HVSs \cite{2008arXiv0808.2469B}. Brown et al. \cite{2006ApJ...640L..35B} estimated that the Galactic halo holds 
$\approx$\,2\,000 HVS in a sphere of 120\,kpc radius. Additionally they introduced the new class of bound HVS \cite{2007ApJ...660..311B} with GRF velocities of more than 275 \kms, which have most likely been ejected by the SMBH in the GC but obviously do not exceed the local Galactic escape velocity. 
The most important attribute of HVSs is, that they provide insight into the environment of the SMBH and its nature as well as  
on the shape and density distribution of the Galactic dark matter halo (\cite{2005ApJ...634..344G},\cite{2006ApJ...653.1194B}).

There are several independent observational results concerning the SMBH environment, which fit in with the existence of HVS: e.g the reason for the high eccentricity of orbits of stars orbiting the Galactic centre may be that they are former companions of HVSs in binaries disrupted by tidal interaction with the SMBH \cite{2006MNRAS.368..221G}. 

A first extrapolation by Yu \& Tremaine \cite{2003ApJ...599.1129Y} showed a HVS formation rate of 10$^{-5}$/yr. 
If the GC hosts a tight {\it binary} of two massive black holes the formation rate of HVSs would be 10 times larger than predicted for a single massive black hole. Therefore once we have found enough HVS to use a statistical approach we could deduce on the structure of the GC.



\begin{figure}[t]
\begin{minipage}{18pc}
\includegraphics[width=14pc]{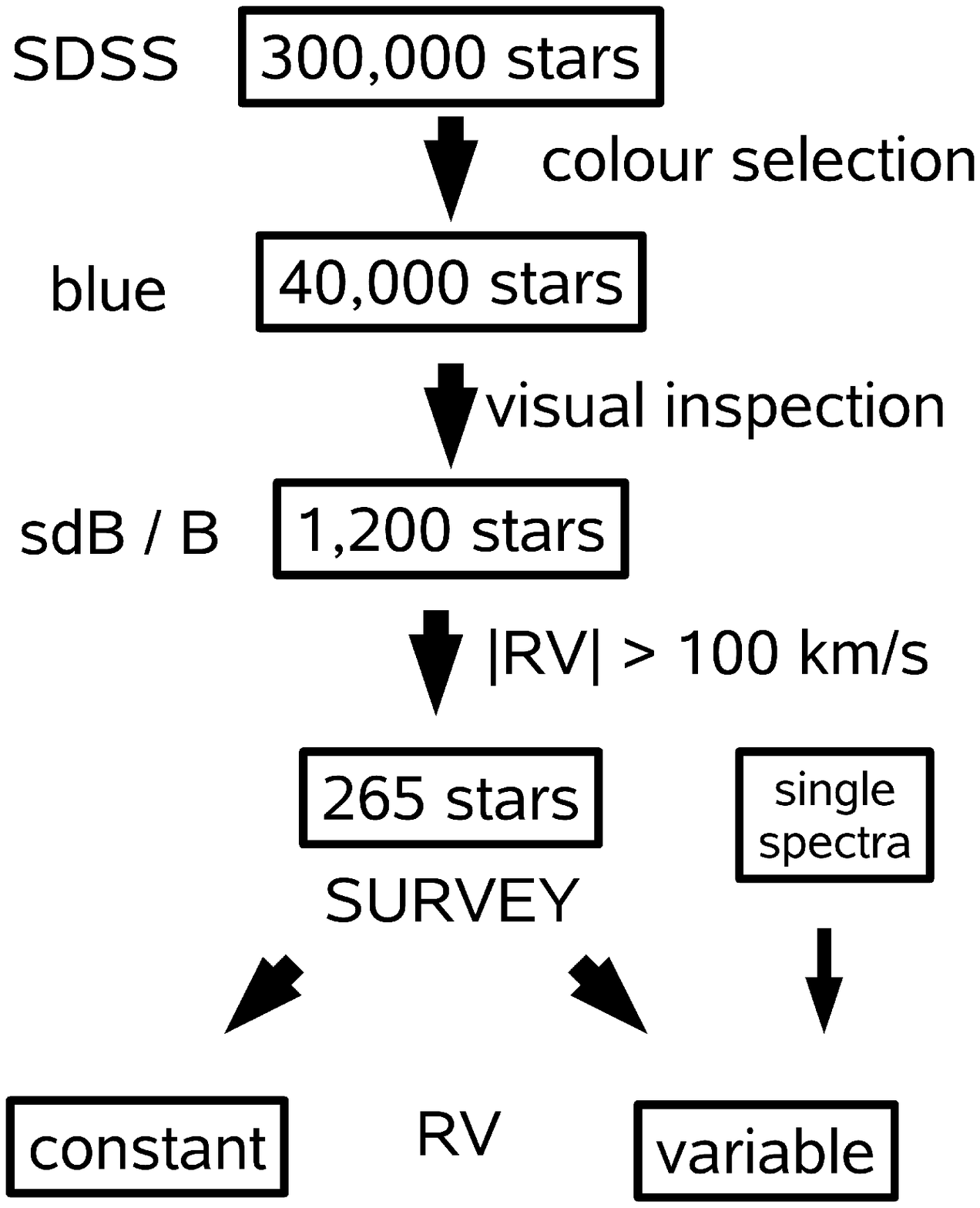}
\caption{\label{label}Target selection method based on Sloan Digital Sky Survey. }
\end{minipage}\hspace{2pc}%
\begin{minipage}{16pc}
\includegraphics[width=20pc]{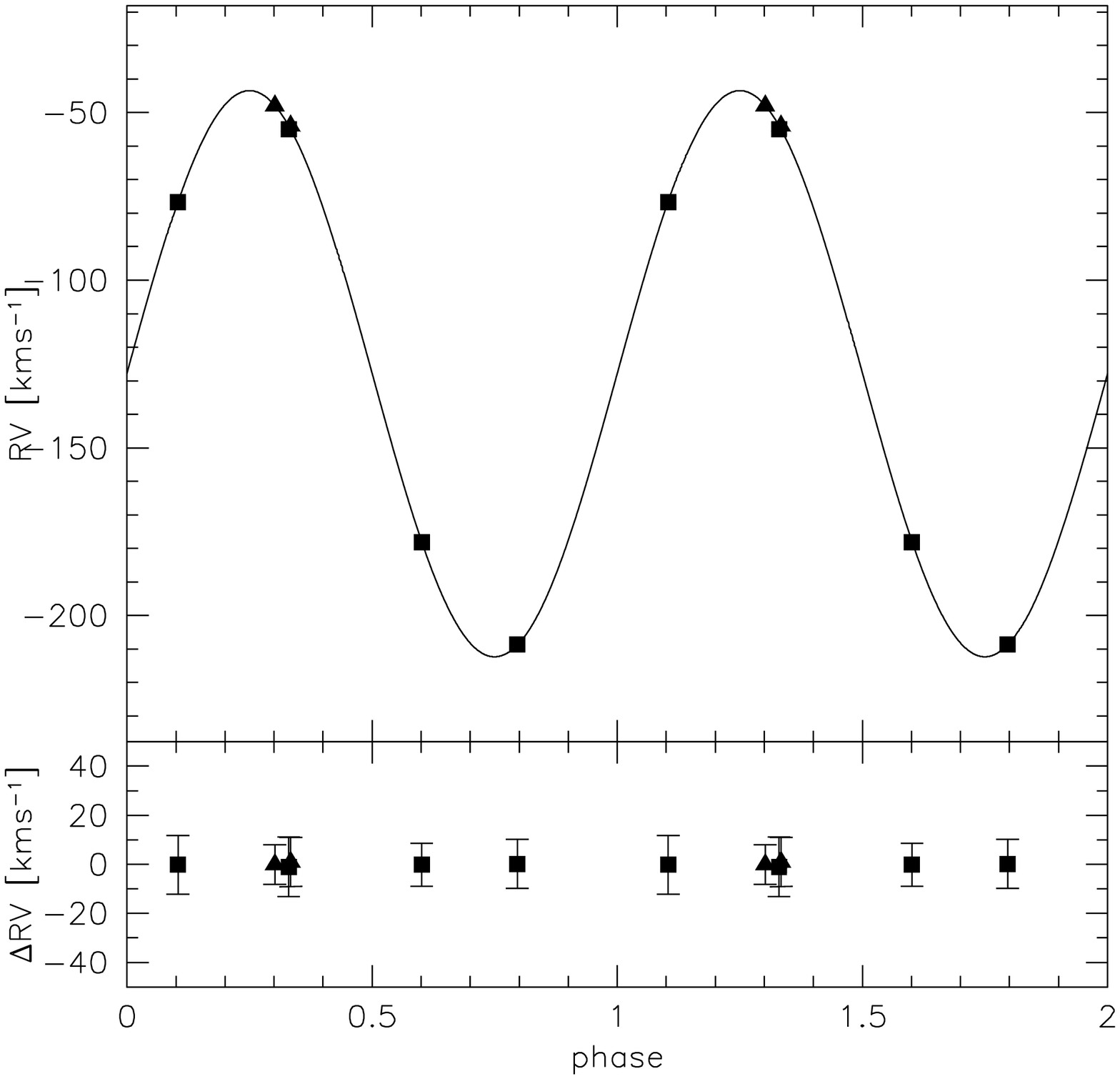}
\caption{\label{label}Radial velocity curve of an sdB binary based on \sdss individual spectra only. Period and amplitude are well determined. 
}
\end{minipage} 
\end{figure}

Strikingly, 13 out of 16 HVS are of late B-type. 
However a discrimination between a main sequence and a 
horizontal branch nature remains dificult, as both branches overlap in the HRD. Only with the help of quantitative spectroscopy one 
can analyse these stars in great detail and draw conclusions as to their nature. Recently one of these stars was discovered to be a chemically peculiar (CP) main sequence star \cite{2008A&A...488L..51P}. Another star seems to be originated in the LMC \cite{2008A&A...480L..37P}, due to its abundance pattern and its location. 

Due to this various implications it is obvious that there is an urgent need to find new HVS and to analyse already
know ones to the highest available precision.

\section{Target Selection}
Unfortunately until now, surveys have only targeted late B-type stars. The only known hot subdwarf was found serendipitously and no systematic search has been undertaken yet. This was one reason for us to start the HYPERMUCHFUSS campaign. 

The enormous \sdss database with its current data release 6 is a perfect starting point for us as the survey is directed towards high Galactic latitudes. The best sky coverage is reached at northern hemisphere at Galactic longitudes close to the GC and the anticenter (Fig. 4). 
To date \sdss is a rich source for stellar astronomy, 
led to various exciting discoveries and has therefore been well evaluated in terms of systematic errors and 
accuracy. 

Our target selection method is shown in Fig. 2. After a colour selection (g-r$<$0.1 and u-g$<$0.4, see Fig.1) and a visual inspection of spectra we select only stars with heliocentric radial velocities higher than 100 \kms.
This reduces the number of interesting stars from 1200 to 265. Only those stars can reach sufficiently high galactocentric velocities to leave the galaxy. 

As all \sdss spectra are a composite of at least three consecutive 15 min exposures, we also extracted the individual spectra for each target, brighter than 18 mag. If the S/N ratio is sufficiently good we are already able to detect rapid RV variability without any additional observations. 
For some of the stars with more images ($\gg3 $spectra) at different epochs one can even find orbital solutions just from these individual spectra (see Fig~3 for an example). 

Our target list consists of a total number of 265 target stars. Fig.~4 shows the target distribution in Galactic coordinates. Although we selected only the fastest stars, the spatial distribution shows remarkable similarities with the general \sdss target sample. Among the targets on the left side in Fig.~4 we find more blue-shifted spectra as the sun approaches them on its disc orbit around the GC, while among the targets on the right side there are more red-shifted spectra for the same obvious reason. 

\begin{figure}[t]
\begin{center}
\includegraphics[scale=0.45]{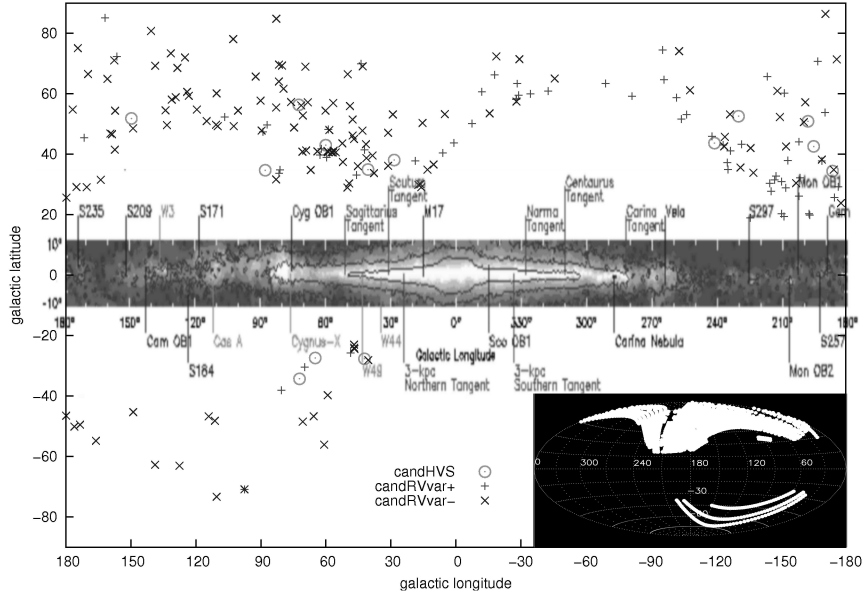}\hspace{2pc}%
\caption{\label{label}Target distribution in galactic coordinates (\textquotedblleft x\textquotedblright:RV$<$0, \textquotedblleft +\textquotedblright:RV$>$0) and the complete \sdss DR6 target distribution (lower right corner).}
\end{center}
\end{figure}

Naturally some spectral misclassification occured, as some of these stars turned out to be other B-type stars rather than sdB stars. But those objects were of interest in their own right.

Finally we calculated the Galactic rest-frame (GRF) velocity and extracted the stars with 
absolute GRF velocity values of more than 275 \kms\, as primary targets for the HVS search. This observational cut is motivated by Brown et al. \cite{2007ApJ...660..311B} to distinguish the bound HVS and runaway halo stars (see above). Henceforth the GRF velocity is used to rank our targets for our survey. Thus we eliminate stars of the disc population and binaries of small RV amplitude, which will mostly harbour companions of relatively low mass. 

\section{Aims and Techniques}
In the primary step our campaign aims at obtaining a second epoch spectra for all our targets (survey), as we use the \sdss spectra as a first epoch measurement. For each star in our campaign there are obviously only two possible results: either the star shows constant RV or we find RV variability. The second step (follow-up) of the campaign then depends highly on what we reveal: 

\begin{figure}[t]
\begin{center}
\includegraphics[scale=1.2]{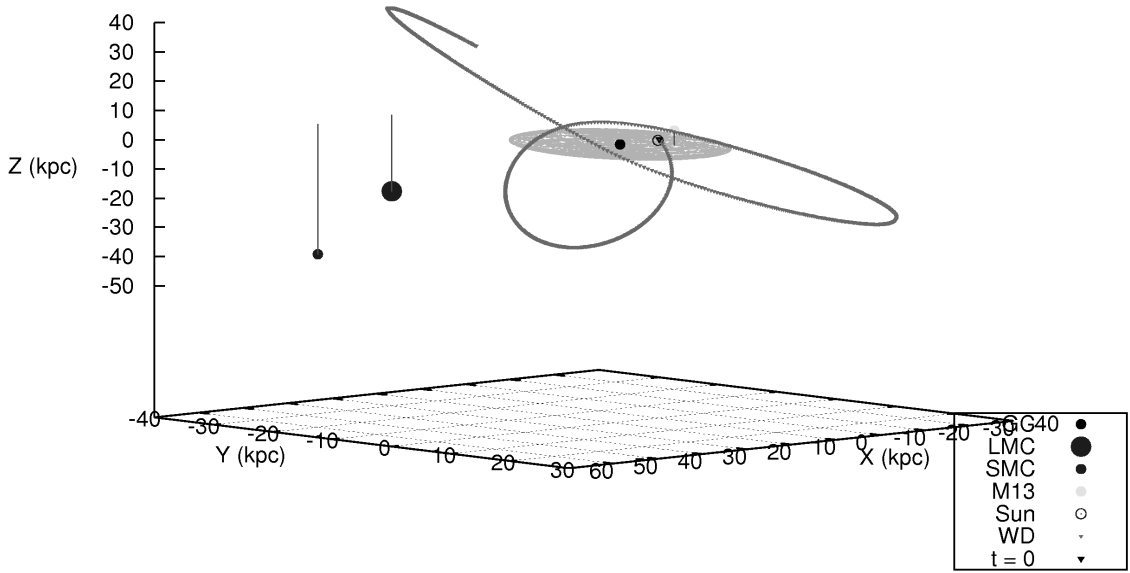}
\caption{\label{label}Trajectory of a halo white dwarf from our sample (known proper motion).}
\end{center}
\end{figure}

in the first case the star is a candidate HVS and a spectroscopic distance determination from high S/N spectra is required to constrain the kinematics of the star. 
In the second case we simply need to get more spectra in order to derive the period and the RV amplitude. 

A $\chi^2$-fitting of synthetic spectra is used to determine the velocity, \teff\ and $\rm{\log{g}}$. To avoid misclassification problems we use our synthetic model spectra for the appropriate stellar types even for the RV fitting. Using the stellar parameters we can infer the stellar masses by applying different evolutionary tracks. 
With the help of the atmospheric parameters, stellar mass and brightness we calculate the spectroscopic distances using the astrophysical fluxes. After that we search the archives for a proper motion measurement of the star or derive a proper motion. It is worth noting that proper motion measurements are not available for any of Brown’s HVSs, making it impossible to determine their trajectories. But with this final step we would be able to calculate a complete trajectory for the star in the galactic gravitational potential \cite{1998MNRAS.294..429D} and probe the galaxy. The future ESO GAIA satellite mission will provide complete proper motion for stars all over the sky as faint as 20~mag.

\section{Preliminary Results}
Our Campaign started in early 2007 and until now we obtained data from the 3.5m telescope at the DSAZ observatory (Calar Alto) in Spain, the ESO NTT and VLT in Chile and the WHT on La Palma. The first step of the campaign consisting of second epoch observations for all possible targets (survey) is currently running and the main part of the data analysis is in progress. 

Until now we observed and analysed 65 out of our 265 targets, i.e. $\approx25\%$. Accordingly 51 of these stars have constant RV within our detection limits and at most 17 stars show indication for binarity, leading to a lower limit for the binary fraction of 27\%. Selecting only the hot subdwarfs we derived an 
even lower limit of only 15\%, which is less than previous estimates (40-70\%) but may be expected if many of our targets belong to population II as the observed binary fraction amongst globular cluster sdBs is very low\cite{2008A&A...480L...1M}. 

In Fig~5 we show the trajectory for a halo white dwarf from our sample, with \teff=16,871\,K and $\rm{\log{g}}=7.38\rm{dex}$. Due to the intrinsic faintness of white dwarfs they are much closer to us than the HVSs, which is confirmed by our method ($d_{WD}\approx400\rm{pc}$). Nevertheless its high GRF velocity combined with a high proper motion value drives the star far out into the halo and indicates population II membership. 

The trajectories shown in Fig. 6 belong to a He-rich subdwarf O star and a late-type B-type star from our sample. According to the atmospheric parameters (T$_{eff}\approx14800K$, $\rm{\log{g}}=4.0dex$) the B-type star is a main sequence star and hence cooler and bigger than we had expected.

\begin{figure}[t]
\begin{center}
\includegraphics[scale=1.2]{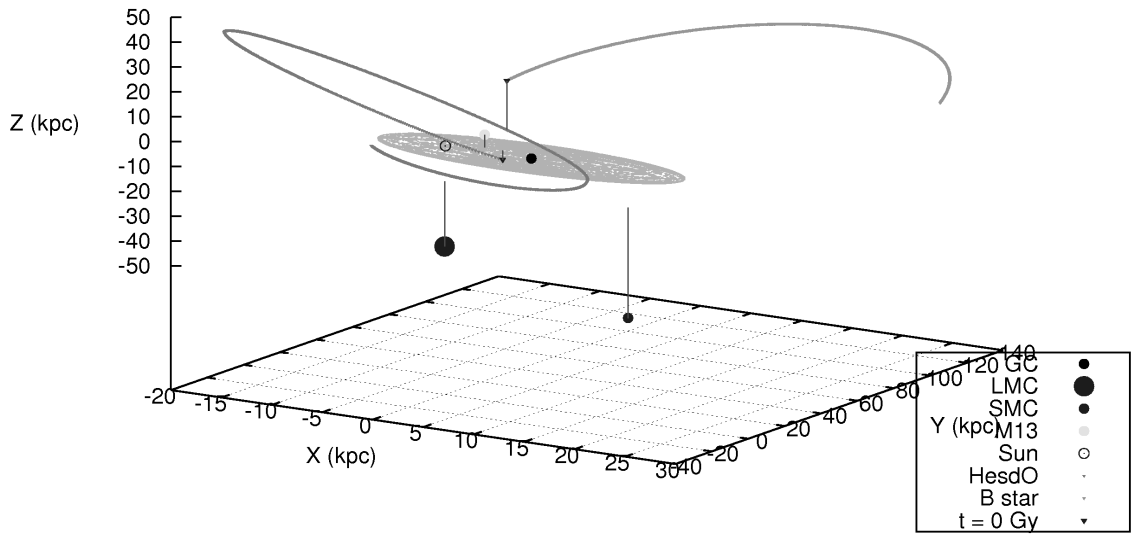}
\caption{\label{label}3D plot of the  trajectory of a He-sdO star(dark grey) and a main sequence B star(light grey) from our sample (unknown proper motion).}
\end{center}
\end{figure}

The GRF at their present locations are -373 \kms\ for the subdwarf O star and +303 \kms\ for the B-type star. If massive, the B-type star is very far away (30kpc) and the Galactic gravitational potential at its present location is fairly small, which cooresponds to a Galacic escape speed of only about 350 \kms. Accordingly this star is also probing far out in the halo. Unfortunately we still do not know its transversal motion. At this location a tiny proper motion already will set this star unbound and turn it into a HVS.

\section{The Search Project}
The HYPERMUCHFUSS campaign targets two different populations, which probe both the GC with its SMBH(s?) and the dark halo as well as a hidden population of black holes. Due to our sophisticated target selection the success of the campaign is guaranteed. Using various state-of-the-art techniques to analyse and interpret the spectra we will derive orbital solutions for the short-period binaries and peform kinematical studies of the HVS. 

We already carried observing runs at DSAZ observatory (Calar Alto) in Spain, ESO NTT \& VLT in Chile and the WHT on La Palma. During the second half of 2008 we plan to complete the first step of the campaign (survey). But the second step of the campaign (follow-up), in which we shall obtain high S/N spectra of the found HVSs and measure precisely the binary orbits, has still to come.

\bibliography{iopart-num}

\end{document}